\documentstyle[prl,cite,amssymb,psfig,twocolumn,aps]{revtex}
\title{Learning, competition and cooperation in simple games}
\author{M. A. R. de Cara, O. Pla and F. Guinea}
\address{
Instituto de Ciencia de Materiales, Consejo Superior de Investigaciones
Cient{\'\i}ficas, Cantoblanco, E-28049 Madrid, Spain.}
\date{\today}
\begin{document}
\draft
\twocolumn[\hsize\textwidth\columnwidth\hsize\csname@twocolumnfalse%
\endcsname

\maketitle

\begin{abstract}
The minority model was introduced to study the competition between agents
with limited information. It has the remarkable feature that, as
the amount of information available increases, the collective gain made
by the agents is reduced. This crowd effect arises from
the fact that only a minority can profit at each moment, while
all agents make their choices using the same input.
We show that the properties of the model change drastically
if the agents make choices based on their individual stories,
keeping all remaining rules unaltered. This variation
reduces the intrinsic frustration of the model, and improves
the tendency towards cooperation and self organization.
We finally study the stable mixing of individual and
collective behavior.
\end{abstract}

]
\narrowtext
The minority game~\cite{CZ97} was first introduced
in the analysis of decision
making by agents with bounded rationality, based
on the \textsl{``El Farol''} bar problem~\cite{A94}. A number of
agents must make a choice between two alternatives. The choice proves
beneficial to a given agent if the total number of agents making
that choice is below a given threshold. The game was formulated
in a precise way by D. Challet and Y.-C. Zhang~\cite{CZ97}.
The bounded rationality of
the agents is modeled by assuming that each agent can only
process information about the outcomes in the $m$ previous
time steps. Given the $2^m$ possible states an agent could afford,
there are $2^{2^m}$ strategies. Each agent has $s$ strategies, taken
at random from the total pool, and for making next decision selects the
best performing one of her own set.
The choice is successful if the agent is in the minority group, 
which means that the
``comfort'' threshold is set at 50\% the total number of agents.
Finally, the agents assign a score to each strategies
at their disposal. The score of the strategies which, at a given time,
have predicted the correct outcome is increased by one point.

The game has by now been extensively studied. Particular emphasis
has been devoted to the mean square deviation of the
number of agents making a given choice,
$\sigma$,
which measures the efficiency of the system.
When the fluctuations are large (larger $\sigma$),
the number of
agents in the majority side (the number of losers) increases.
In this way, the variance
measures the degree of cooperation, or mutual benefit
of the agents. It has been shown that it scales
with $\rho \equiv 2^m/N$~\cite{CZ97,CZ98,Z98,SMR99},
where $N$ is the number of agents
and $2^m$ is the number of
different configurations that the agents are capable
of processing (or
\textsl{states of the world}, see~\cite{CM99}).
When $\rho\gg 1$,
the amount of information available to the agents is so large that they cannot 
manage and exploit it, and agents take decisions like
coin tossing, so that in this limit
$\sigma^2/N \rightarrow 1/4$. In the
opposite limit, $\rho\ll 1$, the set of strategies of different
agents overlap significantly. The agents tend to make 
similar choices, which puts them often in the majority
group. Then $\sigma^2$ scales with $N^2$, instead of $N$.
This regime is highly inefficient from the point of view
of the whole population. The agents manage, however,
to arbitrage away all information in the collective history.
The value of $\sigma$ has a minimum
for intermediate values of $\rho$ which can be appreciated for
not too large values of $s$.
At this minimum, the agents perform better than random, and
some degree of cooperation is established.
This minimum can be understood as a critical point
in an effective spin model with frustrated
interactions and an applied field~\cite{CM99}.

A crucial ingredient in the model is the fact that
all agents act on the same information, irrespective of how it
has been generated. Similar results are obtained when
the histories are replaced by successions of random 
numbers~\cite{C98}, which allows for interesting
analytical analyses~\cite{CGGS99,CM99}.
Evolutionary variations, in which agents with different
number of strategies, $s$, capabilities to analyze the
time series (as given by $m$), or additional adjustable
parameters have also been studied~\cite{CZ98,Jetal98}.
The $\rho\ll 1$ regime leads not only to large values
of $\sigma$ but also to complex distribution probabilities
with a rich structure~\cite{CPG99}.

The model has been used to describe the interactions of
agents competing for scarce resources in different 
contexts~\cite{CZ98,SMR99}. 
However, it is unlikely that the rules by which the agents 
make their choices define a evolutionary stable strategy,
in the sense commonly used in theoretical biology~\cite{MS82}.
The low global gain in the limit $\rho\ll 1$ implies that
alternative rules can easily improve the performance of
the agents. This hypothesis has been verified in different
variations of the minority game as defined above.
Competition between agents with different memories
was first analyzed in~\cite{CZ98}. The rules were extended 
using an additional parameter to improve the chance that the agents
use anticorrelated strategies. The value of this parameter 
was set using an evolution scheme which favors the agent's
performance~\cite{JMJL99}. It has been shown that two populations of agents
with different memories, $m$, perform better than pure
populations taken separately~\cite{JMZH99}. Renewal of the
strategies available to the agents also leads to improvements
in the performance~\cite{LRS99}. In a different context, 
the global gain made by the agents can increase by adding
randomness to the decision making process~\cite{CGGS99}.

We analyze the simplest extension of the model which preserves the basic
structure of the agents' decision process. Each agent has the same
number of strategies, $s$, defined in the usual way, which
process information from the $m$ preceding time intervals.
Unlike in the usual definition of the game, the agents do not
analyze the successions of best choices from the collective point of view,
but respond to the story of the individual choices made by each of
them. Each agent updates the scores of the strategies according to which
strategy, when applied to the individual succession of choices 
made by that agent, leads to a successful outcome.
This is the only difference from the usual case. The processing power
of the agents is exactly the same. 
The model provides a simple way in which agents can avoid
a frustrating situation by ignoring or distorting the information
that has lead them to it.

\begin{figure}
\mbox{\psfig{file=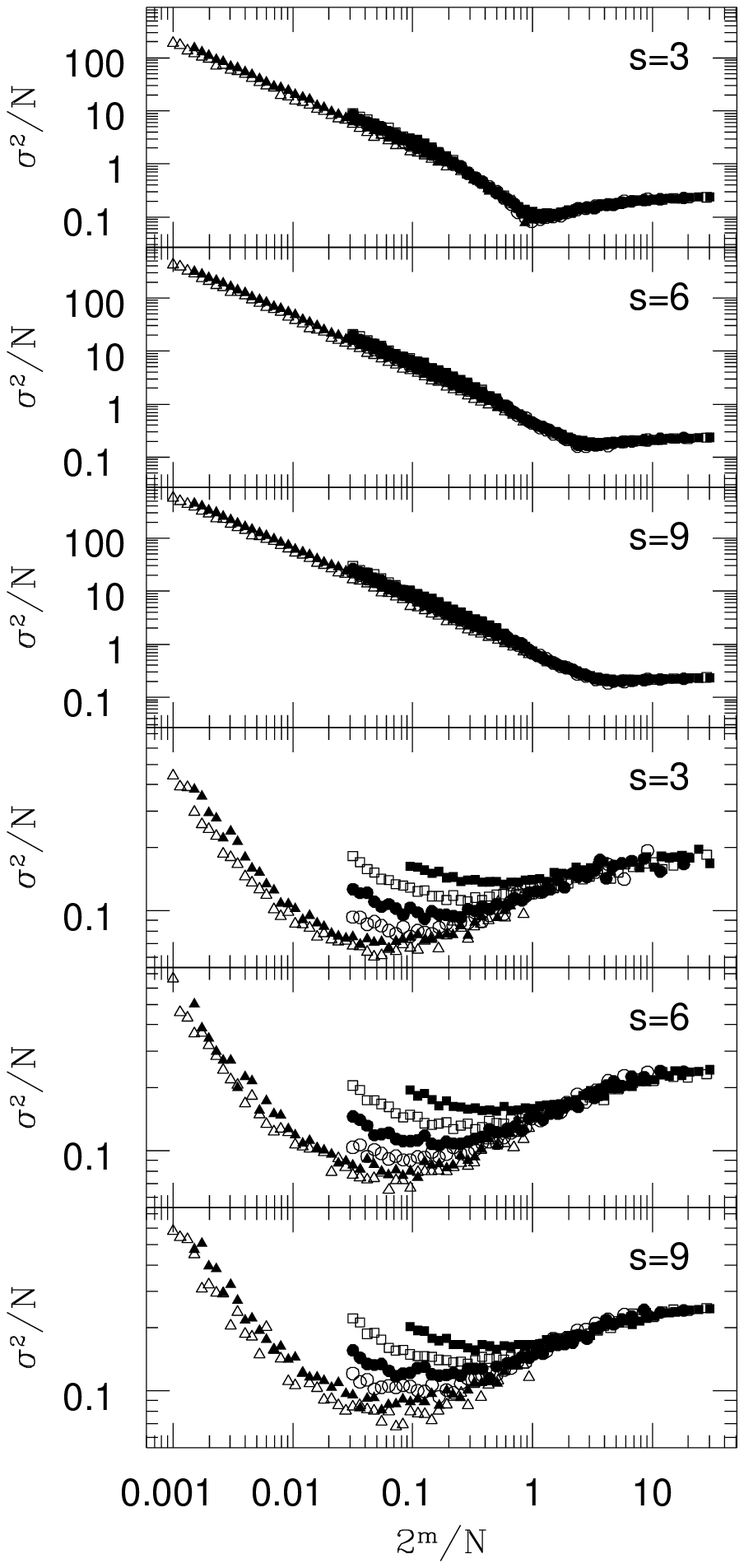,width=\textwidth,angle=0}}
\caption{$\sigma^2/N$ vs $2^m/N$ in the collective (three upper graphs) and
individual (three lower ones) games. Each
point represents the average of 5 independent runs for different values of $N$,
and $m$=4 ($\triangle$), $m$=5 ($\blacktriangle$), $m$=6~($\circ$), $m$=7
($\bullet$), $m$=8~($\Box$), and $m$=9 ($\blacksquare$). For clarity each
value of $s$ is represented in a separate graph.}
\label{fig:comparison}
\end{figure}

We compare the values of the mean square deviations of the 
attendances, $\sigma$, in the present version of the minority game
and the canonical results in fig.~\ref{fig:comparison}.

In the limit when the information available to the agents
is too large, we find
that $\sigma^2 \rightarrow N/4$, the same result as
if the agents made their choices at random, just in the case of high values
of $s$. For case $s=3$ shown in the figure $\sigma^2 \rightarrow N / 5$, and
this value is even lower for $s=2$. $s=1$ is, as in the standard game, highly
sensible to the initial conditions, and averaging over them, gives a dispersion
equal to $N/4$ independently of $\rho$. In the limit
$\rho \rightarrow 0$, the values of $\sigma$ are significantly lower in the 
``individual'' version of the game presented here, and comparable, or lower,
than those found in other extensions of the model. There is a significant
spreading as function of $m$ and $N$,
meaning that the scaling with $\rho$ is not too well satisfied.
The scaling with $\rho$ implicitly assumes that all
possible histories appear
with the same probability
in the collective history~\cite{Z98,C98,CM99}.
In the present version of the model, if the individual stories used by
the agents are replaced by random series,
$\sigma$ takes values close to the random case, irrespective of
the value of $\rho$. Thus, the main hypothesis used to
justify the scaling in the minority game in its usual
form does not hold in this case.

The group which was on the winning side can be
be inferred from the ``comfort'' that
the agent gained after each outcome. This information is used in
updating the score of the strategies, which, however, act on a
different input. As this input is not the same for all agents, they
have no obstacle in following anticorrelated dynamics, even 
when all use similar strategies. The measure of that correlation can be
analyzed explicitly by taking the average Hamming distance
between agents histories~\cite{prepara}.
We have further analyzed this point by calculating the average number
of histories processed by the agents.
The number of histories is always
significantly below that in the canonical model ($P=2^m$),
implying that the system tends to be locked into situations
where agents generate a relatively small number of
possibly anticorrelated individual histories.
This $P$, is also a function of $m$, $N$, and $s$, in
such a way that it decreases monotonically when increasing $N$ and
decreasing $m$.
When $s$ is small the limit for large $m$ and small $N$,
is not $2^m$, but some lower value. This would explain the limit of
$\sigma^2/N\neq 1/4$ when $\rho$ is large discussed above.

\begin{figure}
\mbox{\psfig{file=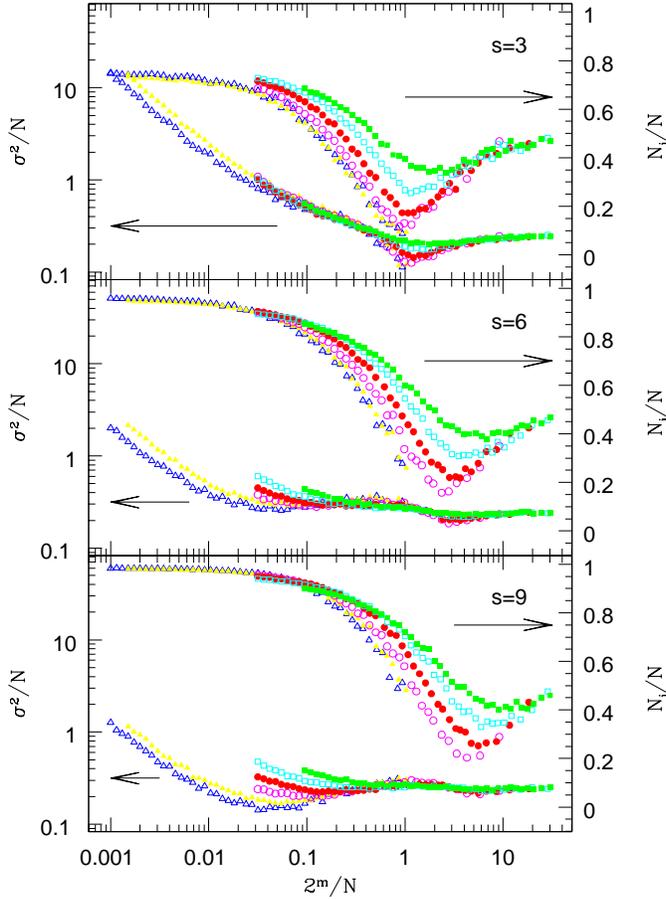,width=0.7\textwidth}}
\caption{Mean square deviation of the attendances in a model
where agents use collective and individual rules (left axis),
and fraction of agents which use an individual rule (right axis).
Different symbols correspond to different choices of $m$
(see fig. \protect\ref{fig:comparison}).}
\label{fig:mixed}
\end{figure}

The present version of the model needs not to define a evolutionary
stable strategy. If there is information available in the 
series of global minority groups, an agent playing
according to the canonical rules will benefit from doing so.
We have analyzed the competition between these two types
of behavior by allowing each agent to have a dual scoring
system for its strategies, following the two set of rules.
Each agent plays the strategy with the highest score at a given
time step. Thus, the population can be divided into those using
collective rules and those using individual rules.
The values of $\sigma$ obtained in this way, and the fraction of
agents using a collective strategy are shown in
fig.~\ref{fig:mixed}.

In the limit when the information available is large,
$\rho \rightarrow \infty$, we recover the random value 
for $\sigma$. Then, both behaviors are indifferent, and the
agents use 50\% of the time each of them.
The fraction of agents which use a collective behavior has
a maximum near the value of $\rho$ for which $\sigma$
has a minimum in the usual version of the model. Finally, the
number of agents using collective rules strongly decreases
as $\rho \rightarrow 0$. In this limit, the preferable
behavior is the individual one outlined here, although a
small fraction of agents using a collective approach
survives. The global efficiency, however, is decreased.
Thus, although a mixed population is the stable situation, the
small fraction of agents which follow collective rules
behave in a parasitic way, lowering the overall gain.

\begin{figure}
\mbox{\psfig{file=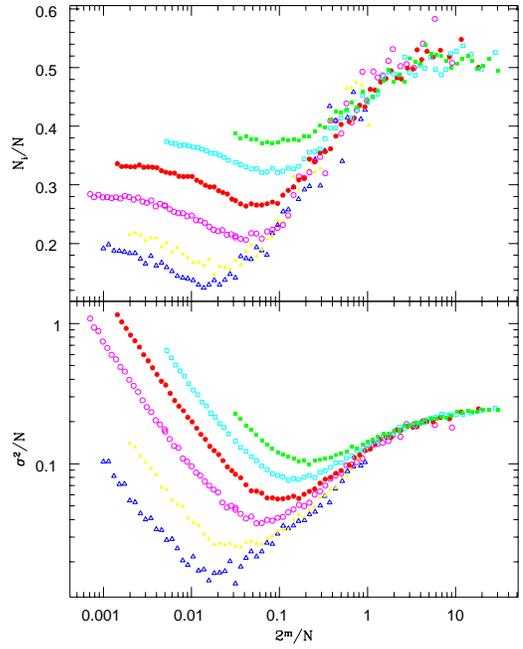,width=0.5\textwidth,angle=0}}
\caption{The same magnitudes drawn in \protect\ref{fig:mixed}, but for the
case $s=1$. Also different symbols correspond to different choices of $m$
(see fig. \protect\ref{fig:comparison}).}
\label{fig:s1}
\end{figure}

The most striking difference with the usual version of the
minority game takes place when only one strategy is
available to each agent, $s = 1$. This case is trivial in 
the minority game, as the agents have no way to learn or to
adapt. The same applies if each agent uses
a purely individual set of rules.
When the agents can use the best of the two
behaviors, the strategy of each agent can be used to process 
two inputs: the collective history of winning sides, or
the succession of prior choices made by that agent.
This is shown in fig.~\ref{fig:s1}.

The global performance of
an hybrid set of agents using both collective and individual rules
is best when $s=1$ for a large range of values of $\rho$.
A qualitative explanation of the adaptability of the agents in this
extreme limit can be obtained by noting that, when a given agent
repeatedly makes an incorrect choice, its individual history
is anticorrelated with the sequence of collective best choices.
Thus, if the strategy at its disposal gives a different outcome
when presented with the two inputs, the agent will tend to give
the opposite answer to that used, unsuccessfully, before. 
There is a self correcting mechanism built into the model,
which tends to prevent very negative performances.
On the other hand, if the agents are locked in into a situation where each
of them obtains about 50\% of the points, 
a stable situation can be achieved, where the agents remain 
anticorrelated by alternating between the two inputs at the
disposal of each of them. This is consistent with the result
that the fraction of agents using collective and individual behavior
is comparable for all values of $\rho$.

In conclusion, we have discussed the simplest extension of the
minority game which preserves the basic parameters of the model.
We show that agents with the same processing power as in the
usual model can perform much better if they use their individual
histories as input, instead of the evolution of the global system.
An evolutionary stable situation arises with agents which can
use both collective and individual rules. The capability of the agents
to adapt and increase the global performance is significantly enhanced,
and herd effects disappear. These emergent features change qualitatively
even the simplest and most trivial version of the minority game, that
in which each agent disposes of a single strategy.

Financial support from the European Union through
grant ERB4061PL970910 and the Spanish Administration
through grant PB97/0875 are gratefully acknowledged.

We acknowledge useful conversations with Damien Challet,
Matteo Marsili and Yi-Cheng Zhang.

\end{document}